\documentclass[conference]{IEEEtran}

\IEEEoverridecommandlockouts
\usepackage{caption}

\usepackage{threeparttable}
\usepackage{tablefootnote}

\usepackage{algorithm}
\usepackage[noend]{algpseudocode}

\usepackage{varwidth}
\usepackage{amsmath}
\usepackage{xcolor} 
\usepackage{hyperref}

\usepackage{graphicx,epstopdf,algpseudocode,caption,url}   
\usepackage{multirow}  
\graphicspath{{./images}}

\usepackage{arydshln} 

\begin{document}

\title{Metaverse in Education: Vision, \\ Opportunities, and Challenges}

\author{Hong Lin$ ^{1,2}$, Shicheng Wan$ ^{3}$, Wensheng Gan$ ^{1,2*}$\thanks{\IEEEauthorrefmark{1}Corresponding author.}, Jiahui Chen$ ^{3}$, Han-Chieh Chao$ ^{4}$ \\ 
	\\
	$ ^{1} $Jinan University, Guangzhou 510632, China\\
	$ ^{2} $Pazhou Lab, Guangzhou 510330, China\\
	$ ^{3} $Guangdong University of Technology, Guangzhou 510006, China\\
    $ ^{4} $National Dong Hwa University, Hualien 974301, Taiwan, R.O.C. \\
	Email: \{lhed9eh0g, scwan1998, wsgan001, csjhchen, hcchao\}@gmail.com
}

\maketitle

\begin{abstract}

Traditional education has been updated with the development of information technology in human history. Within big data and cyber-physical systems, the Metaverse has generated strong interest in various applications (e.g., entertainment, business, and cultural travel) over the last decade. As a novel social work idea, the Metaverse consists of many kinds of technologies, e.g., big data, interaction, artificial intelligence, game design, Internet computing, Internet of Things, and blockchain. It is foreseeable that the usage of Metaverse will contribute to educational development. However, the architectures of the Metaverse in education are not yet mature enough. There are many questions we should address for the Metaverse in education. To this end, this paper aims to provide a systematic literature review of Metaverse in education. This paper is a comprehensive survey of the Metaverse in education, with a focus on current technologies, challenges, opportunities, and future directions. First, we present a brief overview of the Metaverse in education, as well as the motivation behind its integration. Then, we survey some important characteristics for the Metaverse in education, including the personal teaching environment and the personal learning environment. Next, we envisage what variations of this combination will bring to education in the future and discuss their strengths and weaknesses. We also review the state-of-the-art case studies (including technical companies and educational institutions) for Metaverse in education. Finally, we point out several challenges and issues in this promising area.

\end{abstract}


\IEEEpeerreviewmaketitle

\section{Introduction}  \label{sec:introduction}

The concept of Metaverse was first occurred in 1992 and drew attention with its movie Ready Player One \cite{cline2011ready}. Metaverse is a virtual world and closely related to the real life \cite{mystakidis2022metaverse}. It aims to build a digitized world that consists of digital media. The combination between real and virtual worlds represents that virtuality is capable of acting on reality through daily activities and economic life. Studies \cite{kye2021educational,falchuk2018social,huggett2020virtually} suppose Metaverse is ``a 3D-based virtual world in which daily life is done by avatars reflecting the actual individuals''. In other words, he/she will have an easier time finding himself/herself in the Metaverse than in the actual world. In addition, the study \cite{lee2021all} gives another definition of the ``Metaverse'' which is more profound. It is ``a world in which virtual and physical realities interact and co-evolve, and in which social, economic, and cultural activities are carried out to generate value''. In other words, as we described in previous content, the real world and virtual reality are interactions according to the Metaverse instead of two divided worlds. More generally, the Metaverse can also be represented as a world in which daily life and business activities are coordinated \cite{sun2022big,sun2022metaverse,chen2022metaverse}. Some companies and organizations have already been trying to implement Metaverse in several applications, like employee training, student education, and entertainment. The Metaverse is gradually being incorporated into our modern lives. It is necessary and vital to comprehend the Metaverse, and then take full advantage of it.

\begin{figure}[h]
    \centering
    \includegraphics[trim=0 0 0 0,clip,scale=0.28]{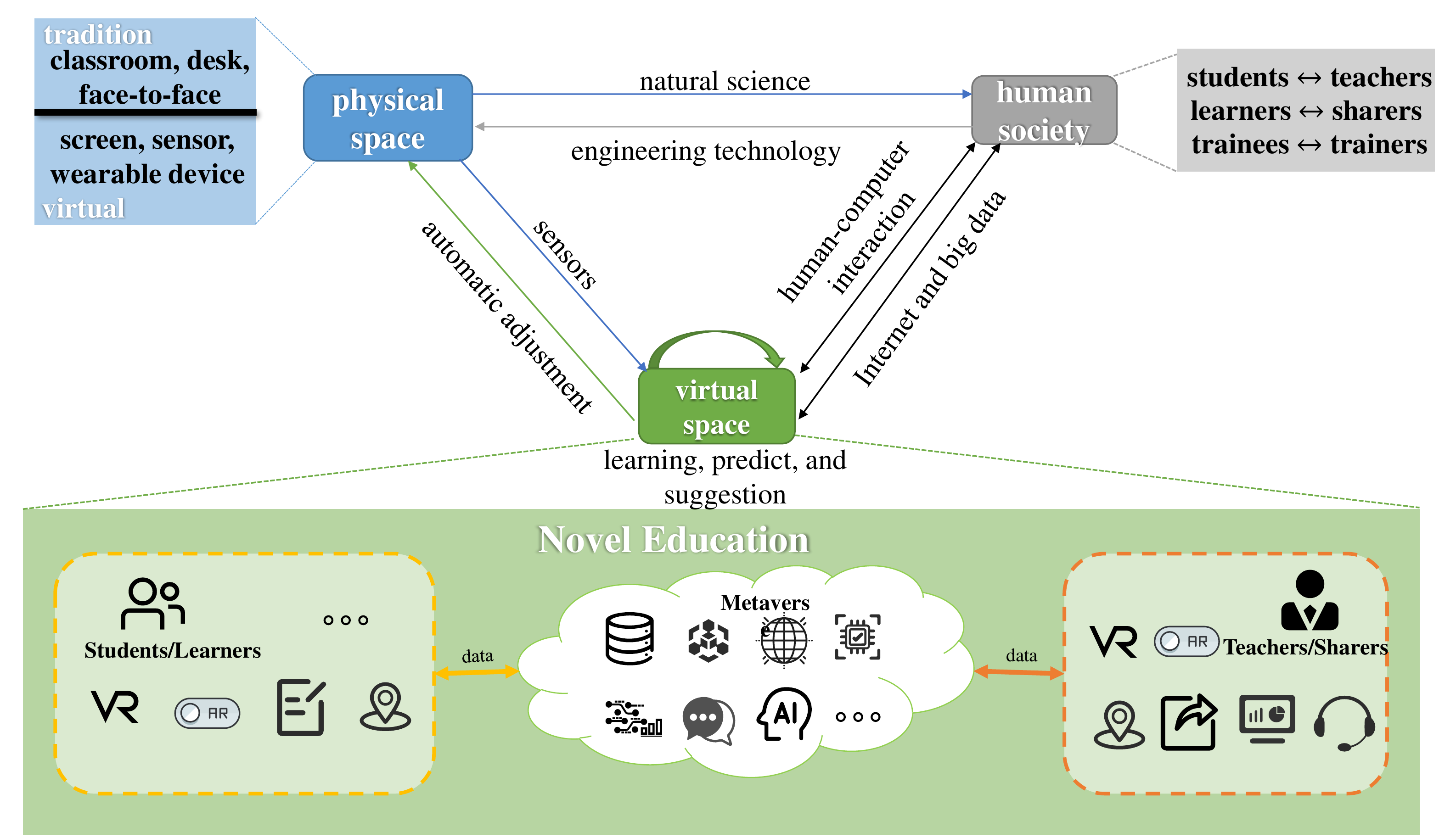}
    \caption{An overview of the Metaverse in education.}
    \label{fig:framework}
\end{figure}

In the long history of human education, word of mouth has been a universal method that mankind has adopted in the wild. Due to the lack of a reliable information carrier, ancestors had to pass down their stories by talking to others, like Homer's Epics\footnote{\url{https://en.wikipedia.org/wiki/Homer}}, Classic of Poetry\footnote{\url{https://en.wikipedia.org/wiki/Classic_of_Poetry}} and so on. Simple, meaningful, inspiring, and easy to remember are the features of these survival stories. However, information distortion inevitably arises during spread, and the listener has to give full play to his or her imagination to understand the story. This case restrains many people's exploration of knowledge to a great extent. Later, the birth of paper allowed people to freely record almost anything. Writing messages on paper is more reliable and convenient than word of mouth, and the paper can save information for a long time. More importantly, the writer can use as many words as possible to clearly describe the scene, which lowers the information entropy to a certain degree. In the past decades, the world has been involved in the Internet age and has developed rapidly. People are no longer satisfied with learning from texts. The novel technologies (e.g., 4G\footnote{\url{https://en.wikipedia.org/wiki/4G}}, streaming media\footnote{\url{https://en.wikipedia.org/wiki/Streaming_media}}, and Bluetooth\footnote{\url{https://en.wikipedia.org/wiki/Bluetooth}}) enrich people's perceptual experiences. We can watch clear photographs or smooth videos on the computer or even on a mobile phone! Through internet technologies, we can easily learn what is happening on the other side of the world. The distance between the world and us is just a network cable. Nowadays, the Metaverse provides a new era for education, including decentralized teaching rooms and immersive studying. The word ``meta'' means first, beginning, important, and consummation. On the one hand, it represents a new beginning; on the other hand, from the perspective of completeness, its connotation already includes not only the virtual and past worlds but also the real and future worlds. At the technical level, it includes many emerging technologies such as big data \cite{sun2022big}, virtual reality (VR) \cite{hu2021virtual}, augmented reality (AR) \cite{maccallum2019teacher}, mixed reality (MR) \cite{nevelsteen2018virtual}, blockchain \cite{zheng2018blockchain,chen2021construction}, digital twin \cite{jones2020characterising}, artificial intelligence (AI) \cite{merabet2021intelligent}, and so on. In short, it is a large integration of human, virtual, and reality across time and space (e.g., cyber-physical space). As shown in Fig. \ref{fig:framework}, the Metaverse education is born at the core of traditional educational activities (e.g., imparting knowledge and learning) and also changes many original things, such as the relationship between teachers and students, the limitation of time and physical space, and so on.

\textbf{Research gap}: Until now, many researchers conducted literature review \cite{ning2021survey,wang2022survey,gadekallu2022blockchain,yang2022fusing} that related to Metaverse in general. Massive basic technologies related to Metaverse have been widely discussed and studied, such as big data \cite{gan2017data,gan2019survey,gan2021survey}, IoT \cite{mozumder2022overview,hajjaji2021big}, virtual reality \cite{miller2020personal}, augmented reality \cite{zhan2020augmented}, mixed reality \cite{siyaev2021towards}, digit twin \cite{liu2021review}, 3D virtual worlds \cite{dionisio20133d}, blockchain \cite{jeon2022blockchain,zhang2020manufacturing,sharples2016blockchain}, 6th-generation mobile communication technology \cite{nguyen20226g}, security and privacy \cite{chen2022metaverse,gan2018privacy}, etc. Though the Metaverse is a buzzword, most people are still unfamiliar with it. Is the Metaverse for education a blessing or a curse? What new functions or concepts are born from the combination of Metaverse and education? Will it cause new conflicts and exacerbate the situation? These questions have remained unanswered.

\textbf{Contributions}: To fill this gap, this paper aims to conduct a systematic literature review of the Metaverse in education. The contributions of this article are as follows.

\begin{itemize}
    \item We introduce a technological framework for the Metaverse in education (called Metaverse education or education Metaverse), which reflects the interlinkages between the Metaverse and smart education.

    \item We elaborate on the description of the characteristics of traditional education, Metaverse, and their combination. We introduce how education, with the help of the Metaverse, will bring changes in the new era. We also discuss how the new education model will contribute to the development of the Metaverse.
    
    \item We provide a detailed survey of the state-of-the-art industry case studies (including companies and universities) for smart education and skill training.
    
    \item We highlight some key challenges and future directions based on our in-depth review, as well as some recommendations for the Metaverse in future education.
\end{itemize}

\textbf{Roadmap}:  In Section \ref{sec:characteristics}, we briefly introduce key characteristics of the Metaverse, traditional education, and their combinations. Then, we discuss how the Metaverse will change education in Section \ref{sec:change}, and put together the state-of-the-art industry case applications in Section \ref{sec:cases}. Furthermore, in Section \ref{sec:challenges}, we provide a comprehensive review of challenges and issues for the evaluation of the Metaverse in education. We conclude this paper with discussions and potential future research in Section \ref{sec:conclusion}.

\section{Characteristics of Metaverse in Education}
\label{sec:characteristics}

\begin{figure}[ht]
    \centering
    \includegraphics[trim=0 0 0 0,clip,scale=0.27]{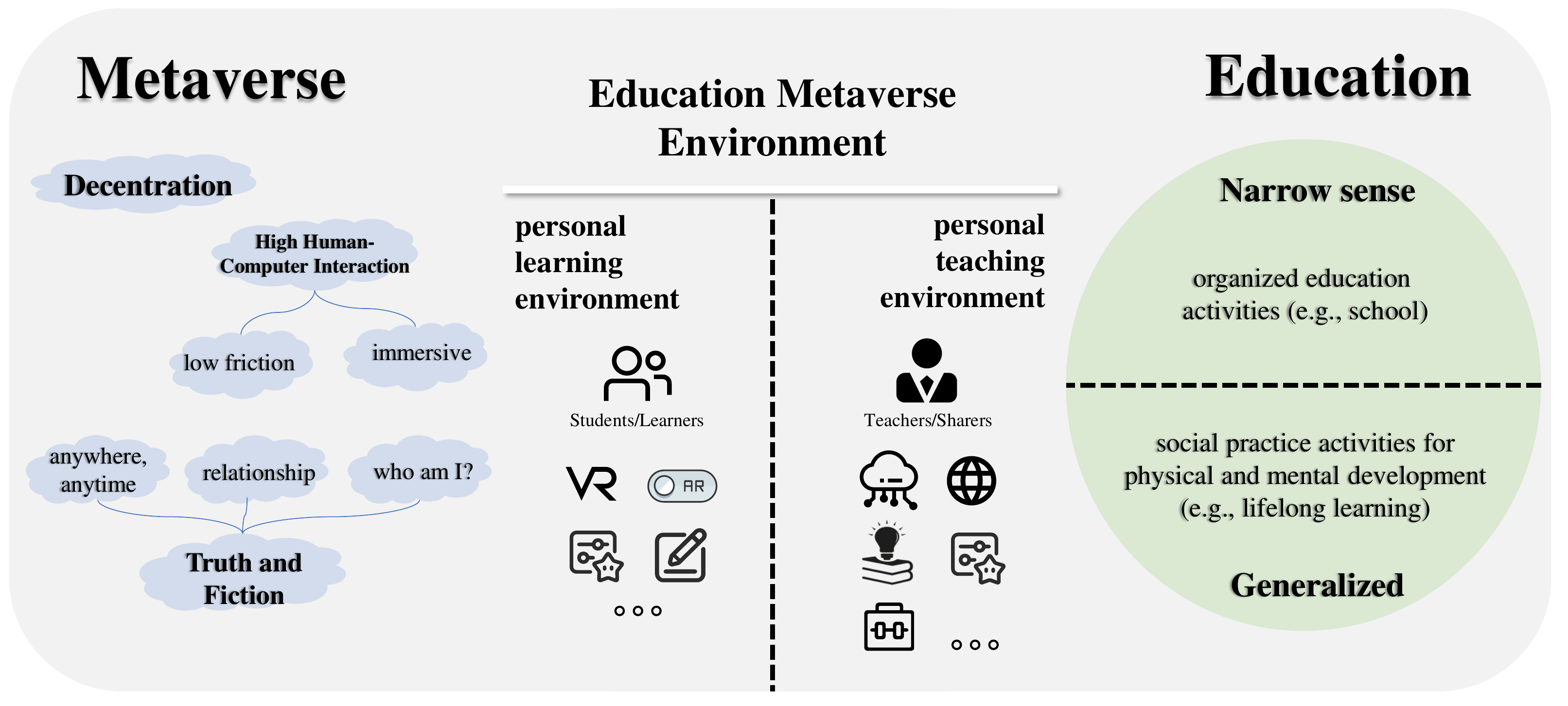}
    \caption{The characteristics of Metaverse and education.}
    \label{fig:characteristics}
\end{figure}

This section introduces some important characteristics of Metaverse, traditional education, and their combination (virtual educational environment), as depicted in Fig. \ref{fig:characteristics}.

\subsection{Characteristics of Metaverse}

As a buzzword, ``Metaverse" is a completely new domain for all people. It is a blueprint for enhancing comprehensive human development. The rapid development of Internet communication techniques and hardware platforms, decentralization, the combination of virtuality and reality, and high human-computer interaction are the conspicuous characteristics of the Metaverse. This is why researchers suppose that the Metaverse in education is possible.

\begin{itemize}
    \item Decentralization originates in the blockchain technology \cite{sharples2016blockchain}. It subverts the traditional social operation mechanism, which means that the Metaverse is no longer created by a specific group of people but by all users' participation and equal co-creation.
    
    \item The Internet of Things (IoT \cite{mozumder2022overview}) means that everything is connected to the Internet. It implements the combination of virtuality and reality, which makes users switch between the virtual and real worlds anytime and anywhere. In a Metaverse world, the identity of a user (who am I?) and social relations in two worlds (what is my relationship with them?) will bring a new way of life.
    
    \item Human-computer interaction \cite{miller2020personal,zhan2020augmented} is the basic function of the Metaverse. Its performance level directly determines the boundary of human sense, and the boundary determines the value of Metaverse to humans.

    \item Digital reality \cite{jones2020characterising,siyaev2021towards} will provide an immersive experience for users. It consists of augmented reality, virtual reality, mixed reality, and 360$^\circ$ video. These technologies help humans get rid of time and space limitations. The immersive technique can greatly deepen understanding and enhance the study experience for learners.
    
    \item The great economic value of Metaverse in education is far greater than that of traditional education. In the Metaverse, everyone is capable of creating various digital products. Besides, transactions among people will be more flexible and convenient (e.g., virtual to virtual and virtual to realistic).
\end{itemize}

\subsection{Characteristics of Traditional Education}

Notice that educational activities have already been changed by the scientific and technological revolution. In a narrow sense, education can be defined as educators (e.g., teachers) conducting specific teaching activities in specific locations (e.g., schools). In a generalized sense, education is the lifelong learning activity of each person (e.g., vocational education, skills training, and enhancing his/her thoughts). During the past decade, with the popularization of the Internet, traditional education has been integrated into Web 2.0. For example, Massive Open Online Courses (MOOCs\footnote{\url{https://www.mooc.org/}}) can realize the online sharing of teaching resources. However, it does not change the core methods because it still relies on content delivery, classrooms, and textbooks \cite{friesen2017textbook}. We suppose it meets many unsolved limitations nowadays, including the lack of engaging teaching content, the low wiliness of student participation, the limited availability of time and space, and the difficulty of concretizing abstract knowledge.

\subsection{Characteristics of Virtual Education}

In the virtual educational environment, according to modern research \cite{starichenko2015interaction, paik2004facilitating}, the educational process is organized, and then it forms a new networked communication space. The educational interaction between students and teachers is supported by a whole management system. The study \cite{yavich2017design} supposes that the virtual educational environment can be roughly divided into two types: the personal teaching environment (PTE) and the personal learning environment (PLE). The core of a PTE is the teachers (or knowledge sharers). With the help of the necessary network services and tools, they share some discipline-specific knowledge on platforms (e.g., blogs, forums, and online communities). Each visitor can freely contribute their comments and download training materials (if the teacher has uploaded them) using cloud tools. PLE has occurred and is continuing to grow. Many online activities or functions (e.g., the creation of virtual space, the collection of information, and the facilitation of communication) are based on precision hardware. It is easy to organize such a virtual educational space on the Internet through cloud technologies \cite{starichenko2015professional} and other web services. Therefore, a PLE should be built and maintained by the learner himself/herself, including all components of educational programs like terminals, communication, and so on. In this type of environment, continuous learning (or called lifelong learning) is feasible for everyone \cite{noguchi2015communities, singh2015global}.

\begin{table*}[ht]
\centering
\caption{Comparison of three types of education}
\renewcommand\arraystretch{1.5}
\label{tab:comparison}
\begin{threeparttable}
\begin{tabular}{|c|c|c|c|}
\hline
                            & \textbf{Traditional Education} & \textbf{Online Education}            & \textbf{Metaverse Education}               \\ \hline
\textbf{Location}               & school                         & school, home                                 & school, home                               \\ \hline
\textbf{Equipment}          & book, pen, blackboard          & computer, mobile phone, tablet       & brain-computer interface, wearable devices \\ \hline
\textbf{Teaching form}      & one-to-many                    & one-to-many, one-to-one              & one-to-many, one-to-one                    \\ \hline
\textbf{Educator}           & teacher                        & knowledge sharer                     & knowledge sharer                           \\ \hline
\textbf{Educated}           & student                        & learner                              & learner                                    \\ \hline
\textbf{Teaching content}   & social and natural science     & interest, social and natural science & customization                              \\ \hline
\textbf{Teaching Purpose}   & personal training              & personal training, enrich lives      & all-round education                        \\ \hline
\textbf{Technology Support} & none                           & Web 2.0                              & Web 3.0                                    \\ \hline
\end{tabular}
\end{threeparttable}
\end{table*}

\begin{figure*}[ht]
    \centering
    \includegraphics[trim=0 0 0 0,clip,scale=0.55]{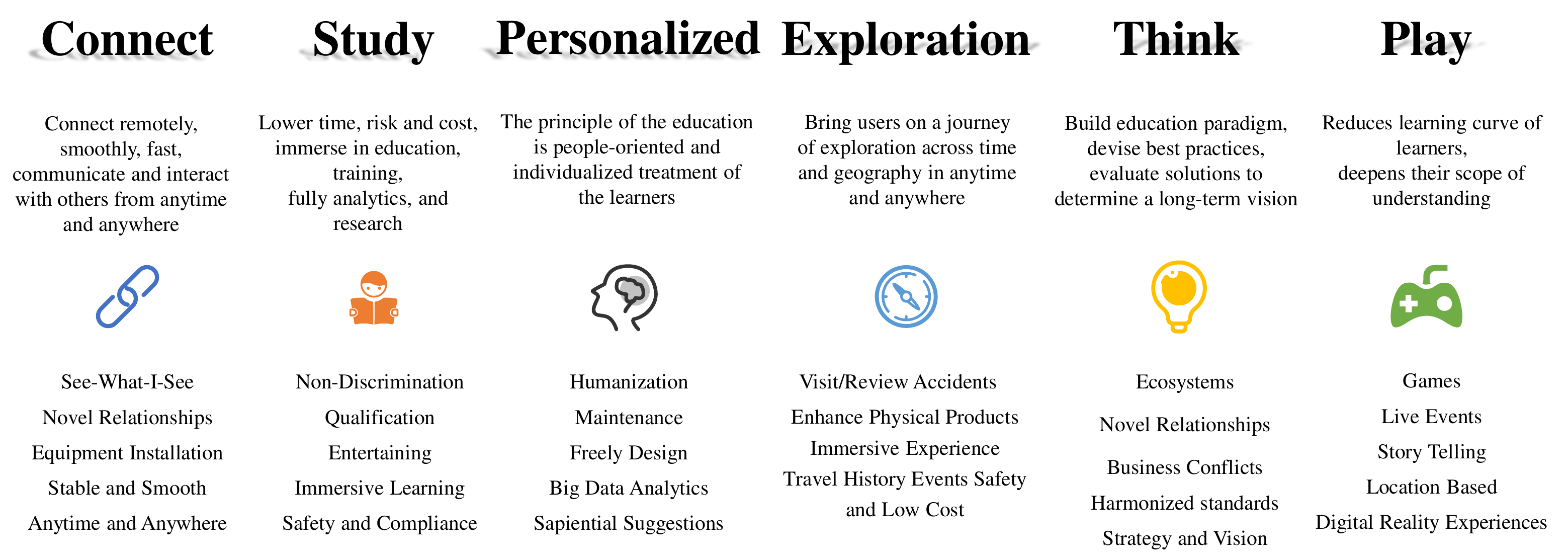}
    \caption{The change that Metaverse brings to education.}
    \label{fig:change}
\end{figure*}

The emergence of new information and communication technologies causes the expansion of educational spaces to shift from offline to online, and this novel shift also updates educational methods and didactic systems. The changes can be roughly divided into two parts. The first is the novel interaction between teachers and students. With the help of virtual educational environments and communications tools, teachers can organize educational activities and collect educational information more conveniently and precisely than before. Big data technologies can strengthen the understanding and memorization of students, and the Metaverse educational system can feed back the requirements of students to teachers in time. More importantly, compared to the traditional teaching model, the virtual environment will ease the seriousness of teachers, and thus students will proactively show interest in interacting with teachers. The second is the new teaching actions. In the virtual educational environment, there are massive templates and instruments for educational information transfer offered to teachers. They can get a detailed analysis of the psychological and educational situations of each student. This will make it easier for teachers to educate students. Based on tools from the virtual educational environment, the method of using a virtual educational environment by a student in learning is undertaken to solve cognitive and educational tasks.

\begin{figure}[h]
    \centering
    \includegraphics[trim=20 0 0 0,clip,scale=0.28]{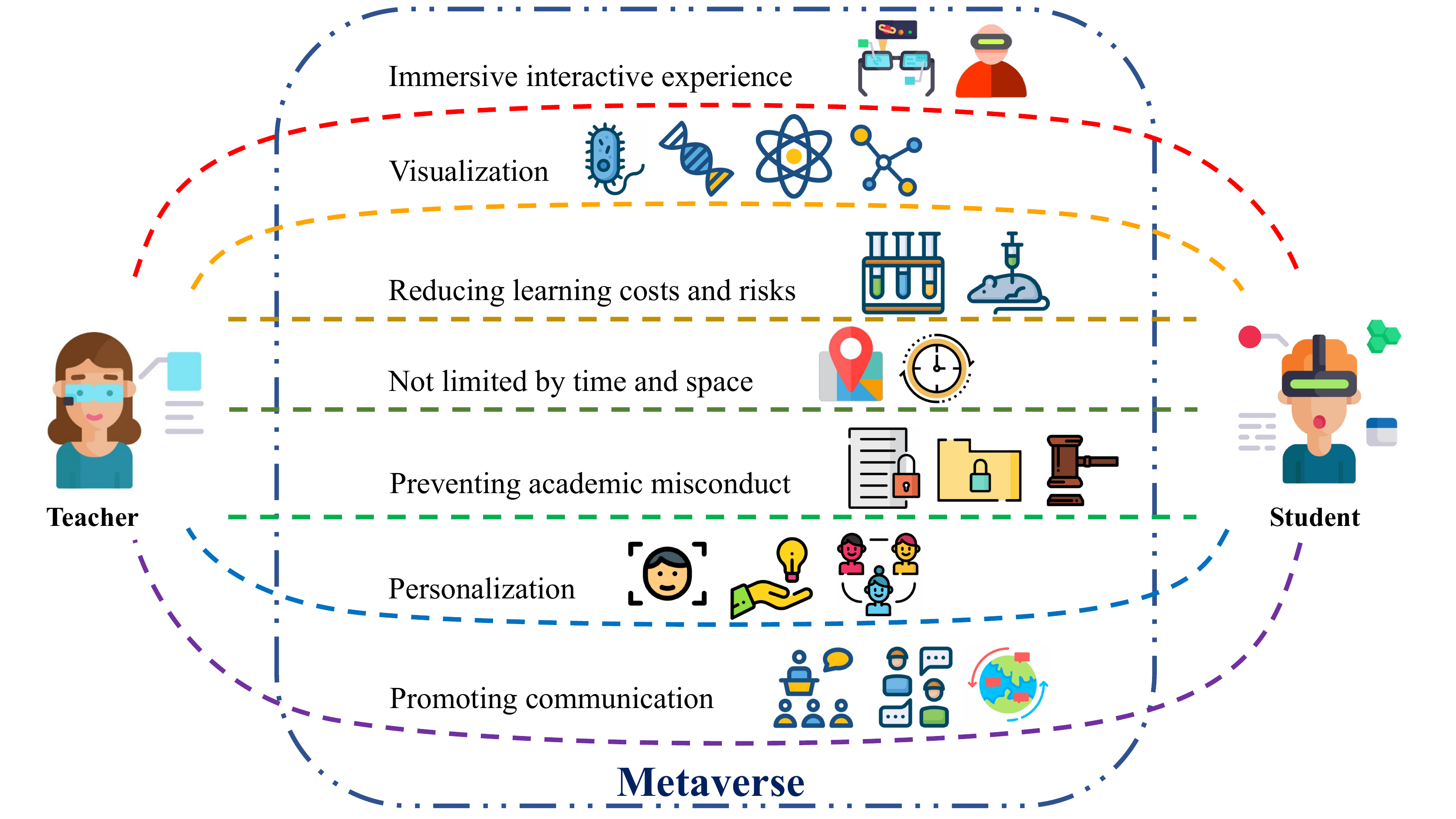}
    \caption{Seven ways that the Metaverse changes education.}
    \label{fig:usage}
\end{figure}

\section{How the Metaverse Changes Education?}
\label{sec:change}

The education system has been developed over centuries and continually adapted to the available techniques. We conduct a simple comparison of three educational models (Table \ref{tab:comparison}). A new revolution in educational models is coming, and we (not only scholars and educators) should embrace and prepare for it. Nowadays, Generation Z is used to accepting online education. In their long lives, the digital world is as important as the real one. Computers, smartphones, and the Internet have surrounded them since they were born. Generation Z are digital natives, and their education must be a challenge (including efficiency and engagement).

Metaverse is an enormous framework that owns many digital features of the future. There are numerous benefits in the Metaverse world, like interaction, authenticity, and portability. As a result, the new educational system has to be readdressed to retain its accessibility and prolong its existence. As shown in Figs. \ref{fig:change} and \ref{fig:usage}, we list some usage of implementing Metaverse in education industry and discuss seven ways the Metaverse can positively impact.

\begin{itemize}
    \item \textbf{Immersive interactive experience}: Metaverse education breaks the limitations of web-based teaching in Web 2.0. Studies \cite{parong2021cognitive,lehikko2021measuring} show that learners will be more enjoyable in a teaching class and thus learn more effectively in a realistic experience (including observation and practice) environment.
    
    \item \textbf{Visualization}: According to digital technologies, the Metaverse can help learners see things that are hardly to contact by their eyes in the real world before, such as molecules or biological cells \cite{thompson2021immersion} in a microscopic view. Besides, it can also simulate ideal conditions in physics, making abstract theories concrete, e.g., Einstein's theory of relativity \cite{georgiou2021learning}.
    
    \item \textbf{Low learning costs and risks}: In general, some classes, like Chemistry and Physics, need to conduct experiments. However, all these experiments can be simulated through digitization, which is a part of Metaverse education. This can save on resource consumption. Similarly, if learners train during high-risk experiments (e.g., with flammable and explosive chemical materials or air crash simulation exercises), the learner's operational risk will be low.
    
    \item \textbf{Unrestricted time and space}: On the one hand, the usage of Metaverse education can be free from the limitation of time. For example, historical events can be recreated and experienced, eliminating the need for students to imagine or watch from books or videos. On the other hand, it breaks geographical restrictions. For instance, students who live in temperate regions want to investigate the environment of tropical regions on the equator. If digital simulations of the tropics have been implemented, students can use the Metaverse to achieve this purpose.

    \item \textbf{Preventing academic misconduct}: Metaverse hunts down academic misconduct through blockchain technology \cite{mohan2019use}. For example, blockchain requires that every generation, release, and flow of information be recorded in a time ledger according to the timestamp. This function can be applied to copyright protection, so that the publication, distribution, and dissemination of academic works can be easily traced and supervised \cite{sharples2016blockchain}. In addition, smart contracts only work when the obligations of the contracted parties are satisfied. When an author submits a manuscript in this system, a new block can be generated, and the transaction information is distributed and stored in the same level block. This case ensures the uniqueness of the author's submission behavior and eliminates academic misconduct (e.g., multiple submissions and multiple releases) to a certain degree.
    
    \item \textbf{Personalization}: Through digital twin generators or simulators, just like the Metahkust\footnote{\url{https://nftevening.com/metahkust-hong-kong-university-metaverse-campus/}} students, learners can design personal avatars according to their preferences, which will make them more confident and engaged in the learning process. Furthermore, after users agree to authorize their personal data, the education system can formulate teaching content and plan courses within the limits of the law.
    
    \item \textbf{Promoting communication}: Due to physical distance, the current online classroom lacks effective interaction and communication \cite{kanematsu2010multilingual}. Learners cannot avoid distraction, and teachers cannot obtain the teaching effect according to the learners' reactions (e.g., facial expressions and body movements) in time. Metaverse allows teachers to create virtual rooms where they can hold internal meetings. At the same time, learners can create study rooms in which they can work collaboratively, study, and socialize freely. According to their avatars, everyone is able to see each other, easily share files, or play games. These features enhance the relationships between learners and teachers (including the friendships of classmates).
\end{itemize}

\begin{table*}[ht]
    \caption{Several representative and productive cases}
    \centering
    \includegraphics[trim=0 0 0 0,clip,scale=0.33]{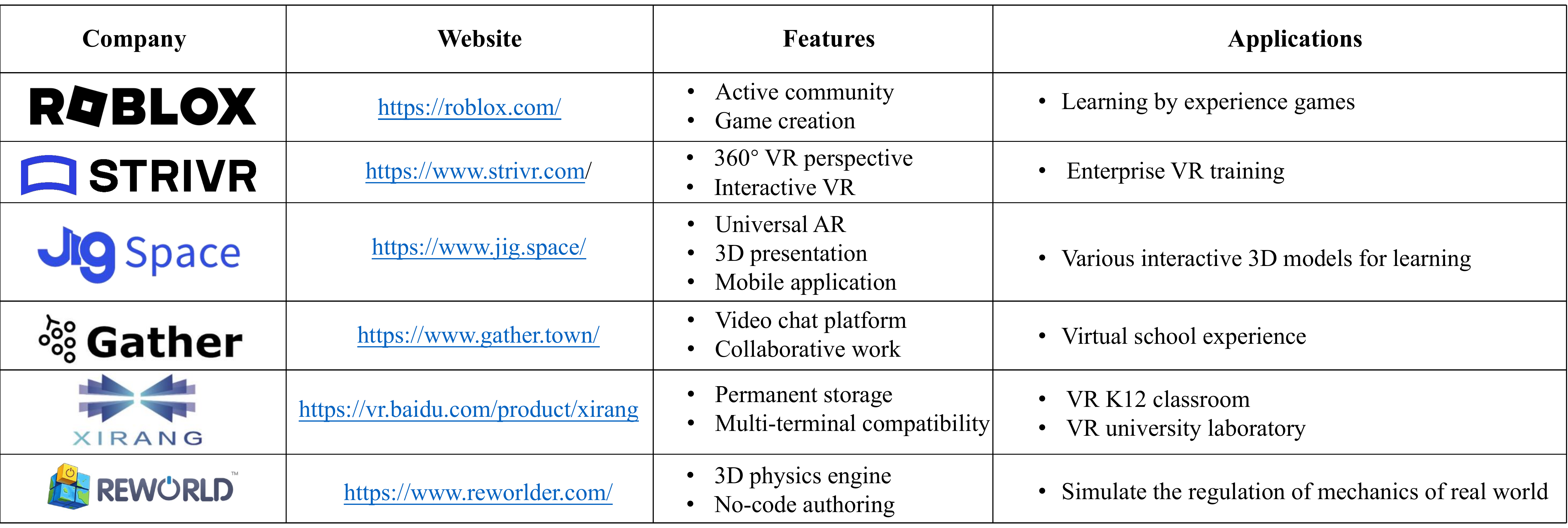}
    \label{tab:products}
\end{table*}

\begin{table*}[ht]
\centering
\caption{Several university's experiences about Metaverse}
\renewcommand\arraystretch{1.5}
\label{tab:university}
\begin{threeparttable}
\begin{tabular}{|c|c|l|}
\hline
\textbf{University}                                                                                        & \textbf{Platform / Tool}                                                                         & \multicolumn{1}{c|}{\textbf{Application}}                                                                                                                                           \\ \hline
Stanford University (America)                                                                     & \begin{tabular}[c]{@{}c@{}}Self development: \\ The “Virtual Human” course\tnote{1} \end{tabular} & \begin{tabular}[c]{@{}l@{}}• Allow all students to break through the space constraints, the\\  “classroom” can be in a museum, laboratory, under the sea, etc.\end{tabular} \\ \hline
Embry-Riddle Aeronautical University (America)                                                    & \begin{tabular}[c]{@{}c@{}}Self development:\\ Extended Reality (XR) Lab\tnote{2} \end{tabular}   & \begin{tabular}[c]{@{}l@{}}• Provide hands-on experiences and augmented learning\\  experiences to serve as supplemental content.\end{tabular}                              \\ \hline
Case Western Reserve University (America)                                                         & Microsoft: Hololens                                                                     & \begin{tabular}[c]{@{}l@{}}• Provide 3D perspective views of parts of the human body.\\ • Enable view perception capabilities.\end{tabular}                                  \\ \hline
\begin{tabular}[c]{@{}c@{}}Hong Kong University of Science\\  and Technology (China)\end{tabular} & \begin{tabular}[c]{@{}c@{}}Self development:\\ MetaHKUST\end{tabular}                   & \begin{tabular}[c]{@{}l@{}}• Provide convenience in notification and administration.\\ • Create your own content freely, such as avatars, NFT.\end{tabular}                  \\ \hline
University of Cincinnati (America)                                                                & \begin{tabular}[c]{@{}c@{}}Self development:\\ UCSIM\tnote{3} \end{tabular}                       & \begin{tabular}[c]{@{}l@{}}• Build a Metaverse learning platform that offers courses in\\  different fields, most notably health care and bioengineering.\end{tabular}      \\ \hline
Soonchunhyang University (South Korea)                                                            & SK telecom: Jump VR                                                                     & \begin{tabular}[c]{@{}l@{}}• Hold the world's first virtual entrance ceremony this year.\end{tabular}                                           \\ \hline
\end{tabular}

\begin{tablenotes}
\item[1] \url{https://stanfordvr.com}
\item[2] \url{https://daytonabeach.erau.edu/about/labs/extended-reality}
\item[3] \url{https://ucsim.uc.edu/}
\end{tablenotes}

\end{threeparttable}
\end{table*}

Therefore, we have a bold vision for the future of Metaverse in education. As we discussed above, the peak of the Metaverse must be decentralized, but the current network ecosystem (e.g., Web 2.0\footnote{\url{https://en.wikipedia.org/wiki/Web_2.0}}) cannot satisfy the requirements of the Metaverse. Fortunately, Web 3.0\footnote{\url{https://en.wikipedia.org/wiki/Semantic_Web}} can be the basic part of Metaverse, which helps machines interpret information like humans. Against this background, we suppose the novel education model will bring some advantages:

\begin{itemize}
    \item \textbf{Low cost}: The Internet of Everything makes it easier to search for information and gather knowledge, which reduces the cost of education for everyone.
    
    \item \textbf{AI search}: The AI search mechanism only displays results that users consider urgent. The search system will automatically mark all different messages and remind users of the relative topics and resources.
    
    \item \textbf{Teaching revolution}: With the help of tools for analyzing big data and educational behaviors, teachers can assign students complex, personalized homework that can help them become more independent.
    
    \item \textbf{Learning}: Students do not need to spend much time collecting and organizing key learning points. Furthermore, they can study or review anytime and anywhere.
    
    \item \textbf{Personal learning network}: The personal learning agent will automatically collect information that is related to the user's learning goal and regularly send them a report.
    
    \item \textbf{Personal education management system}: The management system will make study plans for each user, and the user can change the plan at any time and in any place, based on the current situation.
\end{itemize}

Here, we point out that all the subversive changes that the Metaverse brings to education will also eventually promote the development of the Metaverse itself. On the one hand, many students will benefit from Metaverse-based education, and they will eventually help to promote the Metaverse. On the other hand, the youth who receive education through Metaverse will be more familiar with all aspects of Metaverse. This means that they are more likely to develop the Metaverse.

\section{Industry Case Study of Metaverse Education}
\label{sec:cases}

We are looking forward to the development of the Metaverse in education. The Metaverse has already shown great potential, and there are already some groups focusing on landing Metaverse-based educational projects.

\subsection{Metaverse Education at Tech Companies}

Metaverse education has recently caught the attention of many technology companies. As a significant representative of the Metaverse field, Roblox has unique advantages in Metaverse education. On the one hand, its multiplayer game system provides learners with social interaction experiences. On the other hand, it allows users to create freely, which greatly enriches the content of education scenarios. In November 2021, Roblox announced that it plans to invest more than 10 million dollars in noncommercial organizations and develop three educational games. However, gaming is not the only industry that can benefit from the Metaverse. Strivr is a company that is dedicated to providing VR training to businesses. Its products offer virtual-reality-based immersive learning that engages employees in ``hands-on'' learning opportunities. The results show the learning retention rate of employees increases. More importantly, managers are able to collect unique learning and assessment data to measure training effectiveness and evaluate employee competencies, according to Strivr. In addition, JigSpace is an application that allows users to create 3D models and display their products on an electronic screen. Based on augmented reality technology, it is capable of integrating digital information with the user's real environment in real time. Learners can explore areas of interest in a whole new way through 3D instructional modeling. Here we list some representative and productive cases in Table \ref{tab:products}.

\subsection{Metaverse Education at University}

With the rise of the Metaverse topic, the merits of applying Metaverse to education are increasingly recognized. To test these assumptions, some colleges have begun to experiment with Metaverse education. They all aim at enhancing academic memory and improving educational efficiency through immersive experiences. For example, at Stanford University, professor Jeremy Bailenson was inspired by the novel ``Neuromance'' and thus first set up a course entirely set in virtual reality, called ``Virtual Human''. The ``Virtual Human'' course allows all students to break through the space constraints. According to VR headsets, their ``classroom'' can appear anywhere in the world. Teaching activities are totally done in virtual reality. The ``classroom" can be in a museum, laboratory, under the sea, or even in a ``dangerous" crater. Within immersive experiences, students can experience something that has only existed in their imaginations before. In addition, aviation accidents and safety investigations are one of the main courses at Embry-Riddle Aeronautical University. The headmaster has been trying to build a virtual collision laboratory and make the college a new Metaverse University. In the virtual aircraft control room, students can witness flight accidents, hear the dialogue between pilots and air traffic control centers, and conduct activities such as evaluating emergency response measures, investigating accident scenes, and collecting data. They can enter the accident scene virtually and play as investigators. They can also take pictures, submit their survey records to the professor, and make timely corrections. Besides, Case Western Reserve University School of Medicine is actively applying Metaverse technology to teaching practice. Anatomy students take courses using ``Hololens'', a wearable device based on mixed reality developed by Microsoft Corporation. This pioneering holographic visualization technology provides medical students with 3D perspective views of various parts of the human body and enables view perception capabilities that are difficult to support with other methods. In ``Hololens'', students can freely pan and zoom the hologram to interact with anatomy more naturally and thus improve the efficiency of their anatomy learning. Some works of the university's experience are listed in Table \ref{tab:university}.

After reviewing related products and practices, we believe that the actual experience order of Metaverse education should be as follows:

\begin{itemize}
    \item \textbf{Connection}: This step emphasizes immersive experiences by using simulation exchange (e.g., XR) and digital twin technology.
    
    \item \textbf{Interaction}: This step emphasizes high simulation based on a 3D engine, real-time rendering, and digital twins.
    
    \item \textbf{Creation}: This step is based on the stable operation of the platform, which uses a large amount of data to create content, including objects and spaces.
    
    \item \textbf{Identification}: Identify different users and entities in the virtual world, mainly based on Web 3.0 and blockchain.
    
    \item \textbf{Execution}: Large-scale operations require energy supply, computing power support, and high concurrent transmission capability.
\end{itemize}

\section{Challenges and Issues}
\label{sec:challenges}

\begin{figure}[ht]
    \centering
    \includegraphics[trim=10 0 0 0,clip,scale=0.33]{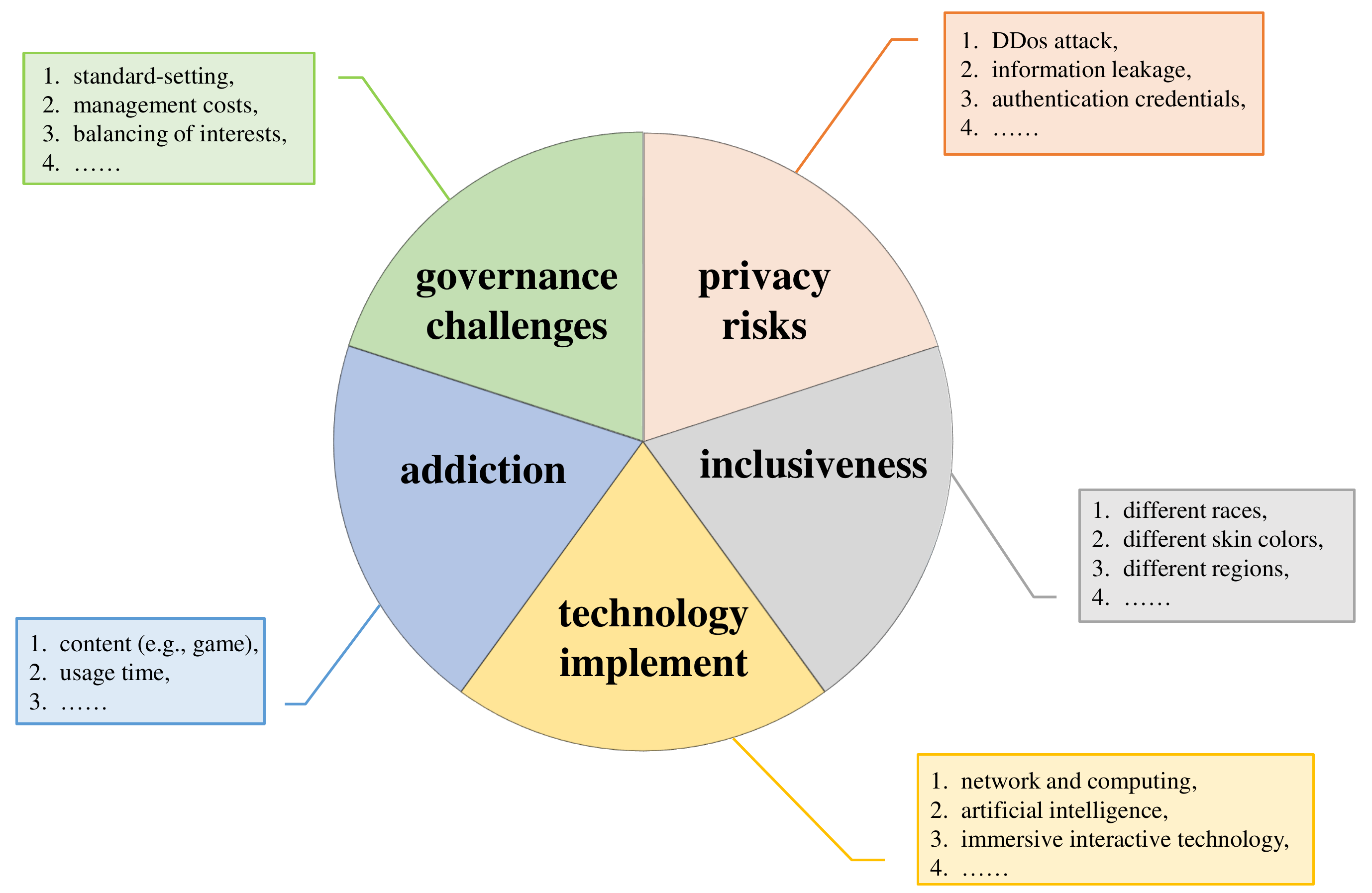}
    \caption{Some challenges and issues of Metaverse in education.}
    \label{fig:challenges}
\end{figure}

As depicted in Fig. \ref{fig:challenges}, we will introduce some challenges and issues in this section. Indeed, there is a massive amount of related content to discuss. Thus, we list five parts (privacy risks, inclusiveness, technology implementation, addiction, and governance challenges) that are urgent to be solved.

\subsection{Technical Implementation}

\subsubsection{Immersive interactive technology}

The realization of the Metaverse is inseparable from immersive interactive technologies, mainly including virtual reality (VR), augmented reality (AR), and mixed reality (AR). In particular, extended reality (XR) is the collective name of the three technologies. We discuss them in detail since they are also educational tools.

\begin{itemize}
    \item \textbf{Virtual reality}: At present, VR is a relatively mature technology in Metaverse education. Previous studies \cite{hu2021virtual,slater2018immersion} have proven that VR has great potential for positive educational results by offering a more engaging environment. However, we lack more in-depth research on this new teaching method. The basic hardware for implementing VR is the head-mounted display (HMD) \cite{choi2017content}, which provides total immersion through a 3D virtual environment to mimic reality. One of the most significant issues with HMD is the lack of visual realism and dynamic interactive realism. It can be concluded that existing technologies that generate VR graphics and display functions are quite limited. We believe this because our human brain is sensitive to change. Even minor details can easily derail the immersive experience. Thus, in a VR world, how to maximize the appearance of reality is an urgent but unsolved challenge.

    \item \textbf{Augmented reality}: Most AR devices are more convenient to wear than VR devices. However, it is difficult to use the same teaching content on different AR devices because many manufacturers do not yet have a uniform standard \cite{marc2021augmented}. Non-portability of content is one of the biggest challenges of AR at present. Besides, the study \cite{rokhsaritalemi2020review} showed that separation of the real and virtual worlds decreases the user's immersion in the educational environment. We need to continually optimize the effective sensing environment of AR and realize the effective fusion technology of virtual information, which depends on the combination of local computing power and cloud computing power, as well as the fusion usage of large broadband and low-latency communication.

    \item \textbf{Mixed reality}: MR requires the coexistence of real and virtual objects, which makes users able to freely interact in real time \cite{hoenig2015mixed}. While MR provides users with a greater sense of immersion than VR or AR, it also imposes many strict technical requirements on its implementation. On the one hand, displaying real objects on a device necessitates high resolution and contrast. On the other hand, the system must accurately track the position of the physical objects because the device needs to give the illusion that the virtual object is in a fixed physical location or attached to a physical item \cite{costanza2009mixed}.
\end{itemize}

\subsubsection{Artificial intelligence}

AI \cite{merabet2021intelligent} is an essential technology to make Metaverse education more effective, such as virtual teaching assistants, language processing for learners from different regions, and learning outcome assessment. For educational AI systems, how to generate virtual teaching assistants suitable for the learners themselves, how to achieve barrier-free communication between people in different languages, and how to reasonably evaluate learners' learning outcomes are three major problems that should be addressed. In addition, the ethical governance of AI decision-making in the Metaverse and preventing the abuse of AI (e.g., utilizing AI code to pass course exams or stealing assignment answers from other learners) in the Metaverse are both non-negligible problems.

\subsubsection{Digital twin}

In the Metaverse, a digital twin \cite{jones2020characterising} is mainly used to digitize real-world objects into the virtual world in real time. In Metaverse education, some experiments require high precision that have strict requirements on the accuracy and real-time performance of digital twins. This means that digital twins should have the ability to identify and fix errors \cite{gadekallu2022blockchain}. Another major challenge of digital twins is data security and privacy when using real-time data to develop digital twin models. See subsection \ref{subsec:privacy} for details.

\subsubsection{Blockchain}

When more and more students accept education in the Metaverse, the current computing speed of blockchain (only three to seven transactions per second \cite{zheng2018blockchain}) may not be able to meet the needs of big data processing. Scalability, which refers to the processing capability of the blockchain network, directly determines whether the blockchain can be combined with Metaverse education. However, when scalability is improved, the number of blockchain nodes participating in the operation will increase, and the probability of a fork in the blockchain will also increase. It means that the decentralization and security of blockchain will be affected, so how to make compromises between these three properties (e.g., the impossible triangle of blockchain) is the dilemma of blockchain \cite{mohanta2019blockchain}. Because of the increased blockchain size, mining costs will rise \cite{park2021promises}. Moreover, the subject of the Metaverse is always human beings, who inevitably make mistakes, especially those who are beginning learners. While blockchain's immutability ensures that data cannot be maliciously tampered with, it also has the limitation of being untraceable. Updating the status by posting new information is not the best strategy, as it takes up extra space and may reveal information \cite{alammary2019blockchain}.

\subsubsection{Network and computing}

The high computational requirements of the Metaverse are also a technical challenge. If education is introduced into the meta-universe, then the simulation and rendering of teaching scenes, the interaction between the teacher and the learner, and human-computer interaction are huge computational quantities, which pose big challenges to the throughput of the network and the computing power of cloud computing.

\subsection{Privacy and Security} \label{subsec:privacy}

Obviously, people will spend more time on the Internet than before, especially with the arrival of the Metaverse. The Metaverse companies will inevitably collect massive amounts of personal privacy information from users to obtain a deep understanding of the user’s thinking and behavior patterns \cite{li2022frequent}. There is no doubt that the collected data will be unprecedented. This causes these companies to have to meet recognition for the protection of personal data and ensure programs are always in place to meet any other requirements (e.g., data regulatory risks). We suppose the weaknesses of the Metaverse education system are focused on the entire life cycle of user data collection, storage, and management, and they should be addressed in the future. We believe that the issue of information security is undeniable. Hence, education in Metaverse should implement the basic properties of information security, including \underline{C}onfidentiality, \underline{I}ntegrity and \underline{A}vailability (CIA) \cite{taherdoost2013definitions}. Here are some scenarios we envision in relation to these three properties in Metaverse education:

\begin{itemize}
    \item \textbf{Confidentiality}: In order to achieve better educational effects, Metaverse education needs to collect identity information or personal emotion tracking \cite{miller2020personal}, the sensitive personal information should be protected in multiple ways. It should be emphasized that the collection of information is not mandatory. If users choose to keep their information at the expense of a better experience, this should be considered reasonable and permissible.
    
    \item \textbf{Integrity}: All information, especially about digital twins, on the educational system cannot be maliciously tampered with, and thus the system should strictly review the uploaded files to avoid penetration (e.g., Trojan horse programs, computer viruses). Otherwise, changing the information may cause damage to the digital twin \cite{far2022applying}. In this regard, combining the Metaverse with a blockchain is a direction worth exploring.
    
    \item \textbf{Availability}: Users are able to enjoy services from the educational system of the Metaverse at any time and any place. Some systems of the Metaverse education can stand up to cyberattacks like DoS and DDoS \cite{umesh2022the, jonathan2022cybersecurity}. 
\end{itemize}

\subsection{Inclusiveness}

The original goal of the Metaverse and education is to encourage more people to engage in them. As a result, it is necessary to create an inclusive virtual environment that takes into account as many different participant requirements as possible. It is challenging but filled with love. For example, affordability is inevitably an issue for poor groups, but they are urgent about accepting education to get a life-changing opportunity. Respecting the needs of special learners, such as the disabled or religious, is often more important than providing a high-quality education \cite{tlili2022metaverse}.

\subsection{Addiction}

As the saying goes, there are two sides to a coin. On the one hand, the higher the quality of immersive interaction, the easier it is for users to indulge in it, eventually leading to ``cyber-syndrome", which means that the physical, social, and mental disorders that affect human beings due to excessive interaction with cyberspace \cite{ning2018cyber}. On the other hand, the XR experience provides learners with numerous visual and auditory stimuli that may increase their cognitive workload \cite{parong2021cognitive,maharg2007simulations}. However, Metaverse education inevitably relies on immersive game technologies (at least now it looks like this). Therefore, how to develop a game that allows learners to achieve learning effects while preventing learners from becoming addicted to them is an important issue for Metaverse education. All in all, Metaverse education should be seen as a tool, not a silver bullet \cite{daniel2022introduction,duzmanska2018can}. Some teaching content that does not need to use Metaverse simulation will not only achieve better teaching results when taught in the real world, but also avoid the addiction.

\subsection{Governance Challenges}

Due to the fact that the moral level of learners varies, it should pay more attention to community governance than to other areas of the Metaverse. Thus, promulgating and improving community standards of conduct to prevent moral problems (e.g., using slang, insulting, bullying, and shaming each other) is necessary \cite{inceoglu2022use}. In addition, since the Metaverse world is larger than Web 2.0, the cost of supervision is also a key problem to be solved. We can learn from the current poor regulation of social media that big tech companies will always put profits before rights or ethics \cite{satvik2022laws}. Therefore, companies that provide Metaverse education platforms can only be operators and should not be absolute regulators. Whether there is a violation or not should be determined by the vast majority of involved users. It should balance the interests of the company and users.

\section{Conclusion} \label{sec:conclusion}

Though we try our best to comprehensively discuss the Metaverse revolution in education, there may still be some technologies or ideas we do not introduce in this paper. We hope that this article will help researchers and practitioners think about potential research directions to pursue while exploring the Metaverse in education. This paper provided an overview of how education meets the Metaverse. Massive research and cases show that combining with Metaverse is a feasible method to achieve relative equality in educational opportunities. Emerging technologies break down many barriers (e.g., space, time, and cost) and thus solve issues that are difficult to address in real life. Metaverse provides outstanding visualization, which cannot be obtained in a traditional classroom. With the rapid evolution of technology, more research efforts are needed to enrich the novel education model with various technologies such as immersive interactive technology (e.g., VR, AR, and MR), network computing, AI, digital twins, and blockchain. This is a new educational environment, and thus we highlighted new educational assessment criteria, governance mechanisms, and study-level testing methods. In addition, it is worth noting that this article also draws attention to the fact that new moral problems need further study. How has the relationship between parents and educational institutions changed? How to safely protect private information? These were briefly discussed.

Standing at the moment when the Metaverse is developing, we can see that the Metaverse and education are mutually accomplished things. Especially in the early stages of the current Metaverse's development, it needs more talents. Education can continuously develop, cultivate, and transport talents for the Metaverse. Thus, in our opinion, the Metaverse and education are closely linked to each other. In the future, how will the application of the Metaverse in the field of education develop, and what potential changes will the field of education need from the Metaverse? With the passage of time, we can look forward to the future of education in the Metaverse era.

\section*{Acknowledgment}

This research was supported in part by the National Natural Science Foundation of China (Nos. 62002136 and 62272196), Natural Science Foundation of Guangdong Province (No. 2022A1515011861), Guangzhou Basic and Applied Basic Research Foundation (No. 202102020277), and the Young Scholar Program of Pazhou Lab (No. PZL2021KF0023). Dr. Wensheng Gan is the corresponding author of this paper.


\bibliographystyle{IEEEtran}
\bibliography{paper}

\end{document}